\documentclass[12pt]{iopart}
\usepackage{graphicx}% Include figure files
\expandafter\let\csname equation*\endcsname\relax

\expandafter\let\csname endequation*\endcsname\relax
\usepackage{physics}
\usepackage{float}

\usepackage{hyperref}

\usepackage{iopams}  
\begin{document}

\title[]{Critical Properties  of Symmetric Nuclear Matter in Low-Density Regime Using Effective-Relativistic Mean Field Formalism}

\author{Vishal Parmar$^1$, M K Sharma$^1$, 
S K Patra$^{2,3}$ }

\address{$^1$ School of Physics and Materials Science, Thapar
Institute of Engineering and Technology, Patiala-147004, India.}
\address{$^2$ Institute of Physics, Bhubaneswar 751005, India}
\address{$^3$ Homi Bhaba National Institute, Training School Complex, Anushakti Nagar, Mumbai 400 085, India}
\ead{physics.vishal01@gmail.com}
\begin{abstract}

The effective field theory motivated relativistic mean-field (E-RMF) formalism is employed to study the equation of state (EoS) for the infinite symmetric nuclear matter at finite temperature using the recently developed forces FSUGarnet, IOPB-I, G3, and the well known NL3 force parameter. The EoS is then used to estimate the critical temperature $T_c$, pressure $P_c$ and density $\rho_c$ of the symmetric nuclear matter for the liquid-gas phase transition. As  $T_c$ is not a constrained parameter in both experiments and theoretical calculations, there is a large uncertainty around its value. Although, the critical parameters are correlated among themselves. It is revealed that vector self-coupling $\zeta_0$ of used forces play determining role in EoS in finite temperature limit. Keeping the incompressibility in acceptable limit i.e. 240$\pm$ 20 MeV, the lower value of $\zeta_0$ gives a better result of $T_c$ when compared to the several experimental data. The critical parameters however show weak correlation with the properties at saturation density at zero temperature. The compressibility factors calculated with these parameters are in agreement with the universal value of liquid-gas systems. Stability conditions are examined along with binodal and spinodal regions. Besides this, the thermodynamic properties like specific heat and latent heat are also worked out. We have carried out detailed consistency check of our calculations using critical exponents and standard scaling laws. All the exponents are well within the theoretical mean-field results.

\end{abstract}

\pacs{21.65.+f, 26.60.+c, 26.60.-c, 25.70.Pq }% PACS, the Physics and Astronomy
                             % Classification Scheme.
%\keywords{symmetric nuclear matter, liquid-gas phase transition, E-RMF, latent heat }%Use showkeys class option if keyword
                              %display desired
\maketitle

%\tableofcontents

\section{Introduction}
One of the vital problems in nuclear physics is to study the nuclear matter and its dependence on density, pressure, temperature, and neutron-proton asymmetry etc. Mahi et al. and  Finn et al. suggested that the liquid-gas phase transition of nuclear matter can be analysed by studying the variation of yield with mass and projectile energy in a multi fragment Proton-Xenon reaction and found that the yield follows a power-law dependence \cite{fermilabgroup1,fermilabgroup2}. This led to several experiments which explored the above mentioned critical nuclear properties in the subsequent years \cite{experiment1,experiment2}. The critical behaviour of nuclear matter is an important feature in medium energy heavy ion induced reactions and therefore numerous theoretical phase-transition predictions have been attempted by several authors over the last two decades. The basis of  these studies is thermodynamics at equilibrium and phase diagram of nuclear matter \cite{theoreticalcalulation1,theoreticalcalulation2,theoreticalcalulation3}. Liquid-gas phase transition  also plays an important role in describing a multi-fragmentation process where intermediate mass fragments (IMFs) are produced in the range of 3$\le$Z$\le$30  \cite{multifragmentation}. It was observed that in  the lowest energy state, nucleons show liquid-like characteristics with density $\rho_0 \approx$ 0.15 fm$^{-3}$ \cite{weleca}. As the temperature increases, the nuclear liquid evaporates and undergoes the liquid-gas phase transition \cite{serot}. Below certain maximum temperature T$_c$, the liquid and gas phase remain in phase coexistence  which terminates at temperature greater than $T_c$ . 

To investigate the liquid-gas phase tranistion in nuclear matter, The Effective-Relativistic Mean Field Theory (E-RMF) has been extended to the case of finite temperature. Quantum chromodynamics (QCD) is the fundamental theory of strong interaction between quarks and gluons that make up hadrons but it can not be used to describe hadron matter as it suffers from nonperturbative properties. Quantum hadrodynamics (QHD) on the other hand is an effective theory of strong interaction at low energy. After the suggestion of D\"urr and Teller \cite{durr}, the first systematic successful model is formed by including only $\sigma$ and $\omega$ mesons called Walecka model \cite{weleca}. This model is not preferred because of its high incompressibility $K_\infty \approx 550$ MeV for the infinite nuclear matter at saturation and its stiff equation of state (EoS). It was  improved by adding the cubic and quartic nonlinearities of the $\sigma$ meson to account for the incompressibility and finite nuclei. The renormalizability of these models constrained the addition of scalar-vector and vector-vector interactions. Subsequently, a series of parameter sets were introduced such as NL1, NL2, NL-SH etc. \cite{nl12} which included $\rho$ meson to take care of the neutron-proton asymmetry and self-couplings of $\sigma$ meson to reduce the incompressibility.  These models still give a stiff EoS at supernormal density.  However, Based on the effective-field theory (EFT) motivated relativistic mean-field (E-RMF) approach, Furnstahl,  Serot,  and  Tang obtained the G1 and G2 parameter sets \cite{g1g2}. Here,  they neglected the idea of renormalizability and included all possible couplings up to the fourth-order  of expansion in the fields which include scalar-vector and vector-vector self-interactions in addition to tensor couplings. It is also well established that the effective Lagrangian  with meson couplings up to fourth order  is a good approximation to predict finite nuclei and nuclear matter observables satisfactorily \cite{g1g2, trunk1,trunk2}. The E-RMF Lagrangian contains all    the    non-renormalizable   couplings   consistent   with   the   underlying symmetries of QCD. The ambiguity in expansion is avoided by using the concept of  naturalness \cite{naturalness}.  The  contributions  of each term in effective Lagrangian is   determined by counting  powers  in  the  expansion  parameters. Field are expanded in terms of nucleons mass to truncate  the  Lagrangian  at  a  given  level  of  accuracy.  For the truncation to be consistent, the coupling constants should exhibit  naturalness  and  none  should  be  arbitrarily  dropped out  to  the  given  order  without  additional  symmetry  arguments \cite{eft}.

The study of nuclear matter EoS at finite temperature has multifold importance. In the laboratory frame, Nuclear Multifregmentation \cite{multifregmentreaction1,multifregmentreaction2}, compound nucleus \cite{cnreaction} , and heavy-ion collision experiments can be used to obtain critical parameters of symmetric nuclear matter. There have been numerous attempts to estimate critical temperature using non-relativistic approaches such as  Thomas-Fermi model \cite{thomasfermi}, Hartree-Fock theory \cite{hartreefock}, Skyrme interaction and Gogny interaction \cite{gognyinteraction} etc. Several calculations have also been done in the relativistic domain using relativistic mean-field framework \cite{ctrmf} and the critical temperature is found in the range $14.2-16.1$ MeV. Apart from the phenomenological models, few  first-principles predictions which uses realistic models of NN interaction  for the liquid-gas phase transition have been carried out using methods such as Finite-temperature self-consistent Green’s function
(SCGF), Brueckner-Hartree-Fock (BHF) formalism \cite{carbone},  chiral effective field theory \cite{corbinian} etc. These microscopic calculations estimate the $T_c$ in the range 14-19 MeV.

At the experimental front, the critical temperature for infinite nuclear matter is measured by extrapolating the data from the multifragmentation reaction on finite nuclei. In these collision experiments, the hot nuclear matter produced has  few hundred nucleons only and therefore such process is not able to describe the infinite nuclear matter.  In contrast to infinite nuclear matter, a hot nucleus shows a considerable reduction in its critical temperature due to effects such as surface tension, Coulomb effects, finite-size effects etc. This reduction is of about 7 MeV leading to the critical temperature of $\approx$ 10 MeV in the mass range A = 50-400  \cite{finitenuclei}. Furthermore, the nature of phase transition is also altered due to strong Coulomb interaction. In this context some authors prefer to define limiting temperature ($T_l$) instead of the critical temperature for the hot nuclei \cite{hotnuclei,tl}. This is the maximum temperature above which Coulomb effects (together with the decrease in surface tension) lead to thermal dissociation of nuclei.  This $T_l$ is discussed experimentally  by studying the plateau in the caloric curve for different types of reactions at different energies \cite{caloriccurve}. Therefore, extending the multifragment reaction for the case of infinite nuclear matter has several limitations arising from finite size, Coulomb, isospin, angular momentum, and secondary decay of excited fragments etc. However the finite nuclei can only be excited up to some limiting temperature

Furthermore, at low density, usually the nuclear matter is not a homogeneous gas of nuclear matter but rather exists in the form of  nuclei of different shapes known as ``nuclear pasta"\cite{pasta} that forms the structure of inner crust of neutron star. The formation of these nuclei of unique shapes lowers the free energy of the system as compared to the corresponding homogeneous phase \cite{avancini}.  However, here we concentrate on ideal symmetric nuclear matter undergoing liquid-gas phase transition.  In the experiments to understand the phase transition in nuclear matter \cite{84, 94, 02, 03, 08, ctnew}, the critical temperature $T_c$ is hardly constrained. There is large uncertainty in the value of $T_c$ among these experiments. Moreover, the model dependence in these experimental calculations arises inevitably. Therefore, the theoretical and experimental calculation of $T_c$ and in general liquid-gas phase transition can not be compared one on one. Instead, to overcome the uncertainty, we look for the dependence of $T_c$ on bulk matter properties such as incompressibility at zero temperature. The correlation of critical parameter among themselves can also be utilised to constraint the related quantities and consequently deducing $T_c$.  To understand this, we use newly developed E-RMF  forces  G3 \cite{g3} ,FSUGarnet \cite{fsu} and IOPB-I \cite{iopb} in the finite temperature limit.  These forces have comparable bulk matter properties at zero temperature although having different numbers of adjustable parameters and their values. Different values of couplings in these newly developed E-RMF sets motivates us to look for the contribution of different terms such as scalar-scalar, scalar-vector terms etc. on the critical values and EoS of SNM at finite temperature. It is to be noted that bulk matter properties are not unique to a particular force as a different combination of these adjustable parameters can give the same bulk matter properties at zero temperature. These facts are used here to study the behaviour of these force in the finite temperature limit and consequently analysing the liquid-gas phase transition qualitatively.

There is a surprising similarity in behaviour near the critical point among systems that are otherwise quite different in nature.  These systems can be partitioned based on their criticality and placed in some ``universal classes".  The systems here liquid-gas system, which belongs to one universal class should have comparable values of critical exponents and compressibility factor keeping with the scaling hypothesis and renormalization \cite{scalinglaw}.  Therefore, a complete statistical study including critical exponents of EoSs derived from different force parameter is necessary to provide a complete qualitative and quantitative understanding of phase transition properties as they all are based on mean-field approximations.  

The paper is organized as follows: In Section \ref{formalism}, we give a brief formalism of the E-RMF model and its extension to finite temperature. In Section \ref{rd}, results and discussions are given. We justify the used forces in Sec. \ref{forces} and the EoS at finite temperature and liquid-gas phase transition is discussed in Sec. \ref{lgpt}. Stability analysis is discussed using binodals in Sec. \ref{StabilityAnalysis}. We have shown the consistency of our calculations using critical exponents along with the well-defined scaling laws or thermodynamical inequalities in Sec \ref{scaling}. We have compared our results with the recent experimental data and various theoretical studies. Finally, the results are summarised in section \ref{summary}.

\section{\label{formalism} Formalism}
 In recent years, the effective field theory motivated relativitic mean field approach is used extensively in  many body nuclear physics problems.  In this theory, nucleons 
oscillate in the mean-field produced by the exchange of mesons and photons.  The  isoscalar-scalar $\sigma$, 
isoscalar-vector $\omega$, isovector-vector $\rho$ and isoovector-scalar $\delta$ mesons interact with Dirac nucleons, i.e. protons 
and neutrons. The biggest advantage of the mean-field theory is that mean-field approximations are thermodynamically
 consistent and it obeys the relevant thermodynamic identities and the virial theorem \cite{covariantcalculation}. 
This provides us with the opportunity to fit our equation of state according to known experimental values at zero 
temperature and then extrapolate it to finite temperature. The basic nucleon-meson E-RMF which involve couplings  of  
$\sigma$, $\omega$, $\rho $ and $\delta $ mesons and the photon with Dirac nucleon up to the fourth-order 
is given as \cite{energydensity,mdelestal}  

\begin{equation}
\label{rmftlagrangian}
\begin{aligned}
&\mathcal{E}=\psi^{\dagger}(i\alpha.\grad+\beta[M-\Phi(r)-\tau_3D(r)]+W(r)+\frac{1}{2}\tau_3R(r)+\frac{1+\tau_3}{2} A(r)\\
&-\frac{i\beta \alpha }{2M}(f_\omega \grad W(r)+\frac{1}{2}f_\rho \tau_3 \grad R(r)))\psi + 
\qty(\frac{1}{2}+\frac{k_3\Phi(r)}{3! M}+\frac{k_4}{4!}\frac{\Phi^2(r)}{M^2})\frac{m^2_s}{g^2_s}\Phi(r)^2\\&
-\frac{\zeta_0}{4!}\frac{1}{g^2_\omega}W(r)^4+\frac{1}{2g^2_s}\qty\Big(1+\alpha_1\frac{\Phi(r)}{M})
(\grad \Phi(r))^2-\frac{1}{2g^2_\omega}\qty\Big(1+\alpha_2\frac{\Phi(r)}{M})(\grad W(r))^2\\&
-\frac{1}{2}\qty\Big(1+\eta_1\frac{\Phi(r)}{M}+\frac{\eta_2}{2}\frac{\Phi^2(r)}{M^2})
\frac{m^2_\omega}{g^2_\omega}W^2(r)-\frac{1}{2e^2}(\grad A^2(r))^2-\frac{1}{2g^2_\rho}(\grad R(r))^2\\&
-\frac{1}{2}\qty\Big(1+\eta_\rho\frac{\Phi(r)}{M})\frac{m^2_\rho}{g^2_\rho}R^2(r)-\Lambda_\omega(R^2(r)W^2(r))
+\frac{1}{2g^2_\delta}(\grad D(r))^2+\frac{1}{2}\frac{m^2_\delta}{g^2_\delta}(D(r))^2.
\end{aligned}
\end{equation}

Here $\Phi(r)$, W(r), R(r), D(r) and A(r) are the fields corresponding to $\sigma$, $\omega$, $\rho$ and 
$\delta $ mesons and photon respectively. The $g_s$, $g_{\omega}$, $g_{\rho}$, $g_{\delta}$ and $\frac{e^2}{4\pi }$ 
are the corresponding coupling constants and $m_s$, $m_{\omega}$, $m_{\rho}$ and $m_{\delta}$ are the 
corresponding masses. Now the  field equations for nucleons and mesons can be solved by applying variational 
principle within the  mean field approximations. The single particle energy of nucleon is obtained by 
solving the Dirac equation constraining the quantum mechanical normalization  as 
$\sum_{\alpha}\psi^{\dagger}\psi_{\alpha}=1$ \cite{diracsolution}.\par
The effective mass of the proton and neutron changes because of  its motion in the mean 
potential generated by the mesons. It is given as 
\begin{equation}
\begin{aligned}
\label{effmass}
&M^*_p& = &M-\Phi(r)-D(r),\\
&M^*_n& = &M-\Phi(r)+D(r).
\end{aligned}
\end{equation}

Now, the E-RMF can be extended to finite temperature. We calculate the grand canonical thermodynamic 
potential $\Omega$ using statistical mechanics as \cite{thermodynamicpotential}
\begin{equation}
\Omega=-k_B\text{T}\ln Z
\end{equation}
and
\begin{equation}
Z=Tr\qty{\exp(-\beta(\vu{H}-\mu\vu{B}))}.
\end{equation}
Where Z is the grand canonical partition function, $\vu{H}$ is the Hamiltonian operator, $\vu{B}$ is 
the  baryon number operator for the  mean field and  $k_B$ is the  Boltzmann constant. The basic relations 
between thermodynamic potential ($\Omega$), entropy (S), chemical potential ($\mu$) and temperature (T) 
are given by
\begin{equation}
\Omega=-pV=E-TS-\mu B
\end{equation}
and 
\begin{equation}
\dd{\Omega}=-S\dd{T}-p\dd{V}-B\dd{\mu}.
\end{equation}
Now the baryon and scalar densities can be calculated by using the ensemble average and are given as :
\begin{equation}
\rho_b=\frac{1}{V}\qty\Big[\pdv{\Omega}{\mu}]=\sum_{\alpha}\frac{\gamma}{(2\pi)^3}\int \dd[3]k[n_k(T)-\bar{n}_k(T)] 
\end{equation}
and 
\begin{equation}
\rho_s=\frac{1}{V}\qty\Big[\pdv  {\Omega}{M}]=\sum_{\alpha}\frac{\gamma}{(2\pi)^3}
\int \dd[3]k\frac{m^*_\alpha}{\sqrt{k^2_\alpha+M^*_\alpha}}[n_k(T)+\bar{n}_k(T)],
\end{equation}
with the $n_k(T)$ and $\bar{n}_k(T)$ are the baryons and antibaryons occupation numbers respectively, which are defined
 by the Fermi distribution function at finite temperature T as\\
\begin{equation}
\label{fdf1}
n_k(T)=\frac{1}{1+\exp\qty(\frac{(E^*(k)-\nu)}{T})}
\end{equation}
and 
\begin{equation}
\label{fdf2}
\bar{n}_k(T)=\frac{1}{1+\exp\qty(\frac{(E^*(k)+\nu)}{T})},
\end{equation}
with E$^*$ as $\sqrt{k^2 + {M^*}^2}$.
The effective chemical potential $\nu$ for proton and neutron in eqs. \ref{fdf1} and \ref{fdf2} are defined as 
\begin{equation}
\begin{aligned}
\label{effcpot}
\nu_p&=& \mu - W(r)+\frac{1}{2}R(r),\\
\nu_n&=& \mu - W(r)-\frac{1}{2}R(r).
\end{aligned}
\end{equation}
The entropy density s ($S=s/\rho_b$) can easily be calculated and has the same form as that for a  
non-interacting gas. It  is given as 
 \begin{equation}
\label{entropyeq}
s_i=-2\sum_{i}\int \frac{\dd^3k}{(2\pi)^3} [n_{k} \ln n_{k} + (1-n_{k})\ln(1-n_{k})+(n_{k}\leftrightarrow \bar n_{k})].
\end{equation}
The energy and pressure expression now can be calculated form energy-momentum tensor T$_{\mu\nu}$ which is  
obtained from Noether's theorem as \cite{weleca} 
\begin{equation}
T_{\mu\nu}=\sum_{i}\partial_\nu \Phi_i \frac{\partial \mathcal{L}}{\partial(\partial^\mu \Phi_i)}-g_{\mu\nu}\mathcal{L}.
\end{equation}
The energy density is calculated using the zeroth component  $\expval{T_{00}}$ and pressure of the system is 
obtained by the third component $\expval{T_{ii}}$ of energy-momentum tensor, i.e. 
\begin{equation}
\begin{aligned}
&E=\sum_{p,n}\frac{\gamma}{(2\pi)^3}\int\dd[3]{k}E^*_{p,n}(k)[n_k(T)+\bar{n}_k(T)]+
\rho W+\qty\Big(\frac{1}{2}+\frac{k_3\Phi}{3! M}+\frac{k_4}{4!}\frac{\Phi^2}{M^2})\frac{m^2_s}{g^2_s}\Phi^2\\
&-\frac{1}{2}\qty\Big(1+\eta_1\frac{\Phi}{M}+\frac{\eta_2}{2}\frac{\Phi^2}{M^2})\frac{m^2_\omega}{g^2_\omega}W^2
-\frac{\zeta_0}{4!}\frac{1}{g^2_\omega}W^4+\frac{1}{2}\rho_3R-\frac{1}{2}\qty\Big(1+
\eta_\rho\frac{\Phi}{M})\frac{m^2_\rho}{g^2_\rho}R^2\\
&-\Lambda_\omega(R^2W^2)
+\frac{1}{2}\frac{m^2_\delta}{g^2_\delta}(D)^2.
\end{aligned}
\end{equation}

\begin{equation}
\begin{aligned}
&P=\sum_{p,n}\frac{\gamma}{3(2\pi)^3}\int\dd[3]{k}\frac{k^2}{E^*_{p,n}(k)}[n_k(T)+\bar{n}_k(T)]-
\qty\Big(\frac{1}{2}+\frac{k_3\Phi}{3! M}+\frac{k_4}{4!}\frac{\Phi^2}{M^2})\frac{m^2_s}{g^2_s}\Phi^2\\
&+\frac{1}{2}\qty\Big(1+\eta_1\frac{\Phi}{M}+\frac{\eta_2}{2}\frac{\Phi^2}{M^2})\frac{m^2_\omega}{g^2_\omega}W^2
+\frac{\zeta_0}{4!}\frac{1}{g^2_\omega}W^4+\frac{1}{2}\qty\Big(1+\eta_\rho\frac{\Phi}{M})\frac{m^2_\rho}{g^2_\rho}R^2\\
&+\Lambda_\omega(R^2W^2)
-\frac{1}{2}\frac{m^2_\delta}{g^2_\delta}(D)^2.
\end{aligned}
\end{equation}

Here $\gamma$ is the spin-isospin degeneracy. $\gamma$ = 2 for pure neutron matter (PNM) and $\gamma$=4 for symmetric nuclear matter (SNM).

After obtaining EoS in pressure and energy, it is straightforward to calculate latent heat using the Clausius-Clapeyron equation
\begin{equation}
\label{classiousclapron}
    L = T \Big(\frac{1}{\rho_g}-\frac{1}{\rho_l}\Big)\Big(\frac{dP}{dT}\Big)_{coex}.
\end{equation}
Here the $\frac{dP}{dT}$ is along the coexistence curve which is determined from the Maxwell construction. It invoke the Gibbs conditions of  phase transition along the isotherms as
\begin{equation}
   \label{chmeicalpotentialeqi}
        \mu_q^{'}(\rho^{'},T) = \mu_q^{''}(\rho^{''},T),\;  (q=n,p),
   \end{equation} 
   \begin{equation}
   \label{pressureeqi}
       P'(\rho',T) = P''(\rho'',T).
   \end{equation}
Where two phases are represented by prime and a double-prime. $\rho_g$ and $\rho_l$ are the coexistence gas and liquid density respectively. Alternatively, the latent heat can be estimated from 
\begin{equation}
    L = T(s_g-s_l), \; (s=S/\rho_b),
\end{equation}   
which is the amount of heat required to take one particle from the ordered phase to disordered phase at a constant temperature, pressure and chemical potential. The related thermodynamic variables such as free energy ($F$), specific heat ($C_V$), isothermal compressibility ($K_T$) can be calculated using standard thermodynamic relations.

\section{\label{rd} Results and Discussions}
\subsection{\label{forces}Force Parameters}
In the present study,  the critical behaviour of the liquid-gas phase transition is investigated using three different E-RMF sets, namely FSUGarnet \cite{fsu}, IOPB-I \cite{iopb}  and G3 \cite{g3}. The results are also compared with NL3 \cite{nl3}  force which is one of the most successful forces used in the finite system calculations. It describes nuclear properties such as quadrupole
deformation and the charge radius exceptionally well for nuclei away from the valley of stability besides nuclei on $\beta$-stability line \cite{abdul}.  The FSUGarnet is a recent parameter set and  IOPB-I is the latest in this series. These both sets are known to reproduce the neutron-skin thickness up to a satisfactory level along with other bulk matter properties.  These give
the adequate addressal of neutron star masses in the lower
(M=2.01 $\pm$ 0.04 M\textsubscript{\(\odot\)} ) and upper limit (M=2.16 $\pm$ 0.03
M\textsubscript{\(\odot\)} ) \cite{iopb}. These both sets have positive $k_3$  and negative $k_4$ corresponding to cubic  and  quartic  terms  arising from the self couplings of the $\sigma$ meson. It is to be noted that a large negative value of $k_4$ leads to the  divergence  of  a  solution  in  the  lighter  mass  region  of  the  periodic table \cite{k4}. The  scalar and vector
cross-couplings $\eta_1$ and $\eta_2$ zero in these sets are zero yet they have  the value of vector self-coupling $\zeta_0$ within the acceptable limit. The positive $k_3$, $k_4$ and small $\zeta_0$ guarantee the agreement with Dirac-Brueckner-Hartree-Fock (DBHF) theory \cite{eft}.\\
The G3 parameter set contains all the mesons and coupling constants in Eq \ref{rmftlagrangian} and therefore describe properties like skin thickness and two-neutron separation energy exceptionally well. The main feature of G3 is that it include the $\delta$ mesons. These are an important degree of freedom for the infinite nuclear matter calculations. Furthermore, due to the finite   $\eta_1$ and $\eta_2$, the G3 set has $\zeta_0$ $\approx$ 1 and positive $k_3$ and $k_4$. All these coupling constants have their specific role in EoS and they, therefore, describe the characteristics of a force.

\begin{figure}[!tbh]
	\centering
		\includegraphics[width=0.6\linewidth]{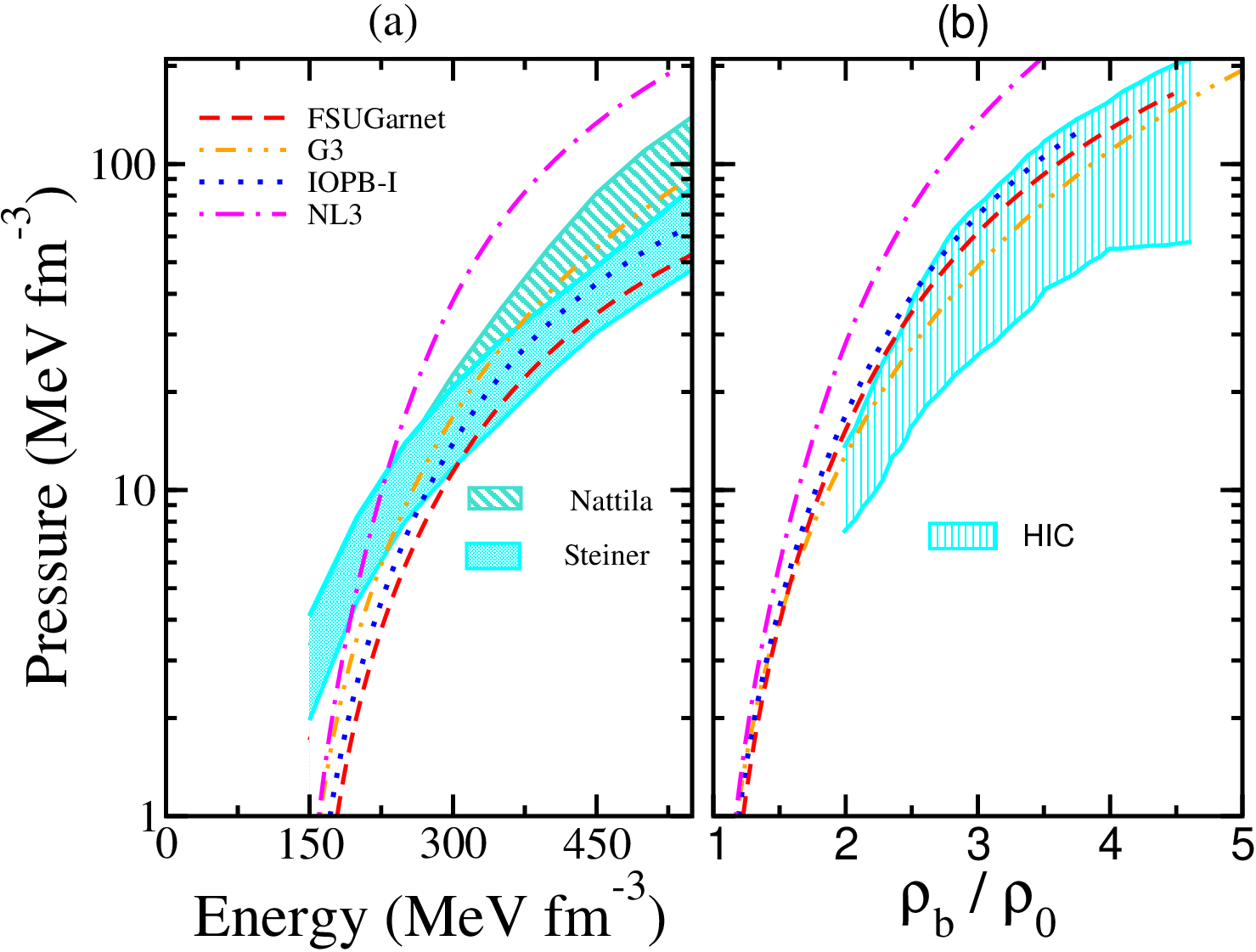}
	\caption{ \scriptsize (a) The pressure versus energy density for NL3, FSUGarnet, IOPB-I and G3 parameter sets 
		at T = 0. The lower shaded 
		region represents the constraint at r$_{ph}$=R with a 95\% confidence limit.  Here r$_{ph}$ and R represents
		photospheric radius and neutron star radius \cite{steiner}. The upper shaded region represents the 
		equation of state for  zero temperature dense matter with a 95 \% confidence limit from  QMC + Model A  
		\cite{nattila}.  (b) The pressure versus baryon density with NL3, FSUGarnet, IOPB-I and G3 sets. The results are compared with experimental data extracted from analyzing EoS after including pressure from asymmetry and density dependence \cite{hic}.}
	\label{eossnm}
\end{figure}
Any effective nuclear force is such that it can explain both finite and infinite nuclear matter.   However, it is difficult using the properties of nuclei as constraints. Consequently, one prefers to extrapolate the systematics of observables such as binding energy, saturation density, compressibility, effective mass etc.  The E-RMF parameter sets considered above have a very narrow range of binding energy, saturation density and effective nucleon mass. They also have compressibility in the  range as determined from
isoscalar giant monopole resonance (ISGMR) \cite{incompressibilityvalue} result. Currently, this value is 240 $\pm$ 20 MeV. When compared with the experimentally derived  EoS's from various sources at zero temperature depicted in Fig \ref{eossnm}, all the parameter sets are in reasonable agreement except the NL3 force due to its slightly larger incompressibility.  This disagreement can be neglected in this work as the domain of density useful for the liquid-gas phase transition is less than 0.15 $fm^{-3}$. Moreover, NL3 set is known to work exceptionally in this density range. Therefore, it becomes essential to investigate the implication of these forces in the finite temperature limit realizing their comparable bulk matter properties at zero temperature.   Since the mean-field theory is thermodynamically consistent, a true  ``universal"  force should not only describe nuclear matter at zero temperature but it should adequately explain the finite temperature properties both qualitatively and quantitatively.  We aim to understand the behaviour of these forces in the context of the liquid-gas phase transition as they have different scalar and vector self and cross-couplings.
.    
\subsection{ \label    {lgpt}Liquid-gas phase transition}
\begin{figure}
	\centering
		\includegraphics[width=0.8\linewidth]{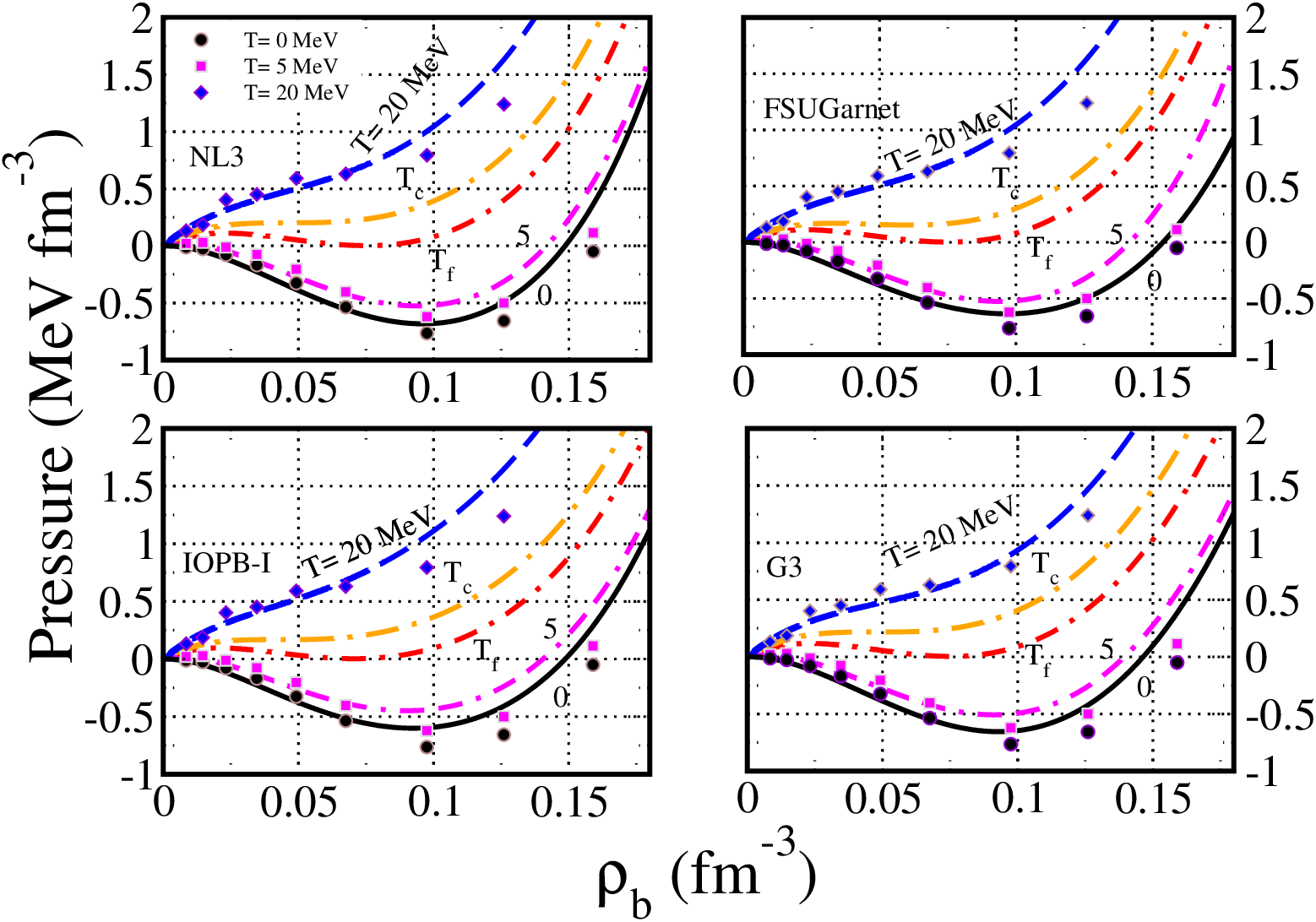} 
	\caption{The pressure versus  baryon density $\rho_b$ at various  temperatures in MeV for the FSUGarnet, G3, IOPB-I and NL3 sets. The dot represents the microscopic calculation by \cite{fp}}
	\label{pvd}
\end{figure}
Now we discuss the implication of these force parameters at finite temperature explicitly near nuclear matter saturation density. The equation of state is shown in Fig \ref{pvd} for the FSUGarnet, G3, IOPB-I and NL3 sets at temperature T= 0, 5, T$_f$, T$_c$, 20 MeV in the pressure vs baryon density phase diagram. Here T$_f$ and T$_c$ stands for flash and critical temperature respectively. It is evident that P-$\rho_b$ isotherms have a quite distinct pocket at a lower temperature. These isotherms show typical short-range Van der Waals like property as the nature of both Van der Waals and Nuclear force is the same. As one move towards higher temperature, the compressibility of nuclear matter decreases and therefore at a particular temperature, we will have a situation where pressure is no more negative i.e. P($\rho$, T$_f$)=0 and consequently $dP/d\rho = 0.0$. This temperature is called the flash temperature and the corresponding density flash density.  It represents the highest temperature at which a self-bound system can exist in hydrostatic equilibrium (P=0). Above this temperature, the nuclear matter is unbound and starts expanding. Further increase in temperature results in the occurrence of an inflation point which is given by \cite{pwang}
\begin{equation}
\frac{\partial p}{\partial \rho}\bigg|_{T_c}=\frac{\partial^2 p}{\partial \rho^2}\bigg|_{T_c}=0.
\end{equation} 
This temperature is called critical temperature $T_c$ after which, the pressure is a monotonically increasing function of density. The corresponding pressure and density are termed critical pressure $P_c$ and critical density $\rho_c$.
At the critical density, the second derivative of free energy ($F=E-TS$) as a function of baryon density is called critical incompressibility. This is given by \cite{yang}
\begin{equation}
   K_c=9\rho_b ^2 \frac{\partial ^2}{\partial \rho_b^2} \frac{F}{\rho_b}\bigg|_{\rho_c}.
\end{equation}

 Table \ref{criticaldata} compiles all these critical values for all the sets. There is a large uncertainty among the several experimental values of critical temperature and it is hardly constrained \cite{84, 94, 02, 03, 08, ctnew}. This uncertainty mainly arises because all these experiments are performed for fragmentation reaction on finite nuclei and then extrapolated to evaluate the $T_c$ for the infinite matter. These experimental calculations are model dependent. Besides, various effects like finite size,  small time scale $\approx 10^{-(22-23)}$s in multi fragment reaction which makes it hard to study thermodynamic equilibrium, Coulomb interaction, isospin, angular momentum etc., also add to this uncertainty. Therefore, it is not very wise to theoretically address the quantitative nature of the nuclear matter phase transition using any mean-field calculation. Although one can realise based on recent experiments \cite{03,08, ctnew} that critical temperature should be greater than 15 MeV. 
	
\begin{table}[t]
		\caption{The calculated critical temperature $T_c$ (in MeV), critical pressure $P_c$ 
			(in MeV fm$^{-3}$) and critical density $\rho_c$ (in fm$^{-3}$) with the respective binding energy  
			(in MeV) at the critical point  along with the flash temperature $T_f$ and flash density $\rho_f$ with FSUGarnet, IOPB-I, G3 and NL3 parameter sets for symmetric nuclear
			matter. The experimental data are given wherever available. The corresponding incompressibility, effective mass at critical density and compressibility factor $C_f$ and critical incompressibility $K_c$ is also shown. } 
     	    \resizebox{\textwidth}{!}
     	    {\begin{tabular}{ c  c c c c  c  c  c  c  c c c c }
			\hline
			\hline
			& $T_c$  & $P_c$ & $\rho_c$   & $\mathcal{E}/\rho_b$ at $T_c$ &T$_f$&$\rho_f$&$K_\infty$& $M^*_c/M$&$C_f$ & $K_c$   \\ 
			& MeV & MeV/fm$^3$ & fm$^{-3}$  & MeV& MeV& fm$^{-3}$&MeV &&& MeV  \\ 
			[0.5ex] 
			\hline
			FSUGarnet  & 13.80 & 0.171 & 0.043 & -9.61&11.3&0.071&229.50&0.850&0.282&-68.35 \\  
			IOPB-I  & 13.75 & 0.167 & 0.042 & -8.80&11.2&0.071&222.65&0.864&0.277&-69.95 \\ 
			G3 & 15.30 &  0.218 & 0.049 & -6.68&12.1&0.075&243.96&0.879&0.292&-82.91 \\ 
			NL3 & 14.60 & 0.202 & 0.046 & -8.38&11.8&0.070&271.38&0.846&0.276&-77.85 \\
			G1 \cite{g1g2}& 14.30&0.187&0.046&-8.24&11.5&0.075&215.00&0.877&0.285&-72.03\\
			G2 \cite{g1g2}  &14.30&0.184&0.043&-8.04&11.8&0.080&215.00&0.879&0.299&-78.77\\
			NL1 \cite{nlct} & 13.74 & 0.164& 0.041 &9.71&11.2&0.070&211.70&0.872&0.290&-71.909\\
			NL2 \cite{nlct} & 18.63 & 0.361 & 0.056 &-3.47&14.3&0.085&399.20&0.861&0.345&-111.03\\
			NL-SH \cite{nlct} & 15.96 & 0.264 & 0.052& -6.70&12.7&0.080&355.36&0.846&0.315&-90.097\\
			FP (Microscopic) \cite{fp}& 17.5    $\pm$ 1.00 & - &- &-&-&-&240.00&-&-&-\\ 
			Exp. 1 (1984) \cite{84}& 12.0 $\pm$ 0.20 &- &- &- &-&-&-&-&-&- \\ 
			Exp. 2 (1994) \cite{94}& 13.1 $\pm$ 0.60 &- &- &- &-&-&-&-&-&- \\  
			Exp. 3 (2002) \cite{02}& 16.6 $\pm$ 0.86 &- &- &- &-&-&-&-&-&- \\  
			Exp. 4 (2003) \cite{03}& 20.0 $\pm$ 3.00 &- &- &- &-&-&-&-&-&- \\  
			Exp. 5 (2008) \cite{08}& 19.5 $\pm$ 1.20 &- &- &- &-&-&-&-&-&-\\
			& 16.5 $\pm$ 1.00 &- &- &  &-&-&-&-&- \\  
			Exp. 6 (2013) \cite{ctnew} & 17.9 $\pm$ 0.40 & 0.31 $\pm 0.07$ & 0.06 $\pm$ 0.01 &-&-&-&-&-&0.288&-\\[1ex]
			\hline
			\hline
		\end{tabular}} 
		\label{criticaldata}
	\end{table}
Theoretically, there are two sets of RMF parameters. One, those have incompressibility in the accepted range as predicted by ISGMR result i.e. 240 $\pm$ 20 MeV. These parameters, which include the FSUGarnet, IOPB-I and G3 forces,  predict the critical temperature less than 15 MeV. One exception being G3 which estimates $T_c$ = 15.3 MeV. These sets are extensively used for Neutron star calculation where the nuclear matter is at high density. The others set, although estimate large $T_c$ but have large incompressibility and therefore are not used in nuclear astrophysical applications.  Our calculations with the set FSUGarnet, IOPB-I, G3 along with the NL3 force are in reasonable agreement with the first set of forces.

The critical parameters obtained using mentioned E-RMF sets shows some striking behaviour in terms of their correlations which otherwise are very difficult to establish analytically. $T_c$ and $P_c$ have the strongest correlation with Pearson's coefficient 0.991 with p-value of 0.009. Here  p-value  represents the significance level of correlation coefficient. This is in consistency with  classical van der Waals (VDW) gas property ; $T_c/P_c$ =8b, where b arises due to the repulsive interaction \cite{wanderwaal}. A similar trend is observed in $T_cT_f$,$T_cK_c$,$P_cK_c$ and $\rho_cK_c$ with p-value less then 0.05. These correlations are in consistent with the standard analytical equation relating the $K_c$ and the ratio $P_c/\rho_c$ given by \cite{yang}
\begin{equation}
    K_c + 18\frac{P_c }{\rho_c} = 0,
\end{equation}
Therefore, these correlation suggests the strong relationship of critical parameter among themselves. They become important as constraining $T_c$ is difficult  and a direct comparison with experimental data is not suitable. If however, extrapolating the finite nuclei experiments to infinite matter was possible accurately and critical temperature in these experiments could be measured precisely, one can easily get other finite temperature properties based on established correlations.\\
\begin{figure}
	\centering
		\includegraphics[width=12cm,height=6cm]{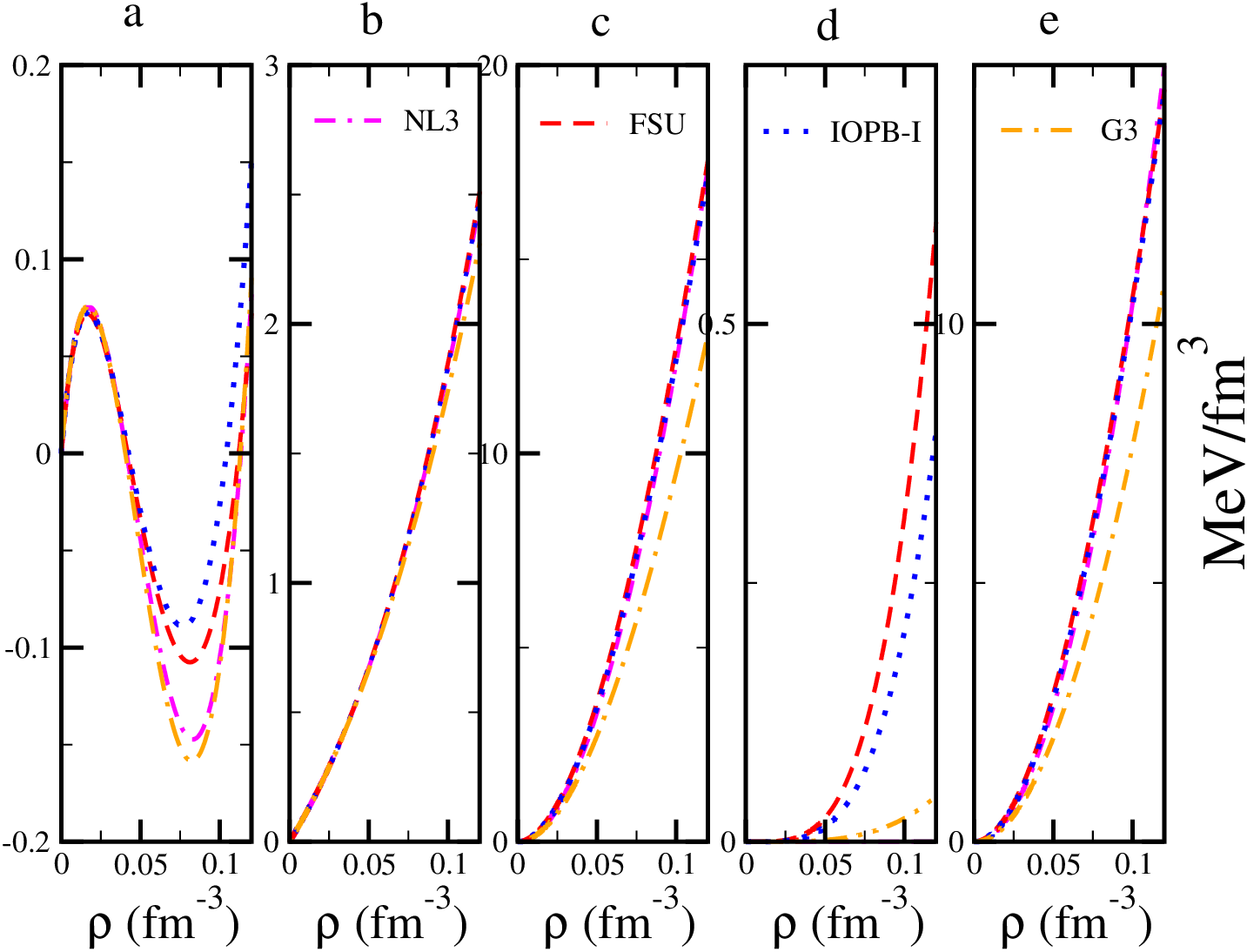} 
	\caption{The contribution arising from the different meson interaction on total pressure at T=10 MeV is shown for the different parameter sets. (a) Total Pressure, contribution due to (b)  kinetic part (c) scalar-self interaction (d) vector-self interaction and (e) scalar-vector cross interaction}
	\label{comparison}
\end{figure}
On the other hand, weak correlation of critical parameters with that of saturation properties at zero temperature is also interesting. Naively, one could argue that the binding energy at saturation should be related to critical temperature. Instead, there is no such correlation between the binding energy $e_0$ and $T_c$ or $T_f$.  Similarly these models also do not satisfy the empirical relations \cite{rhocempirical} between saturation density  $\rho_0$ and $\rho_c$ or $\rho_f$. We observe that the $T_c$ has a spread of 1.55 MeV while the saturation energy and saturation density lie within 0.27 MeV and 0.005 $fm^{-3}$ respectively. It indicates that the properties at saturation have no dictation over the position of critical points. Therefore, any  relation, relating properties at saturation and finite temperature  can not be generalised.
It should also be noted that each of the above force i.e. FSUGarnet, IOPB-I, G3 and NL3 have different adjustable parameters. One can not make any prediction directly by these adjustable parameters. In other words, none of the scalar and vector couplings (self and cross) ($k_3, k_4, \zeta_0$) individually has a direct correlation with the critical parameters. However, accept the NL3 force, a greater value of $\zeta_0$ gives smaller $\rho_c$ and force with greater incompressibility estimates the larger $T_c$ .  A close investigation of the contribution arising from scalar and vector channel suggests very strongly that these forces only differ in contribution from vector self-coupling term and therefore, $\zeta_0$ becomes an important quantity in finite temperature EoS. This is shown in Figure \ref{comparison} where contribution from different interaction in total pressure is shown at T=10 MeV. The only major difference is  in graph (d) which represents the vector-self interaction term in the pressure. The effect of small value of $\zeta_0$ in G3 and absence of it in NL3 is visible. The NL3 consequently have a large negative $k_4$ which is a possible reason for it to not show the common characteristic of other three forces. A soft contribution of vector-self interaction in total pressure therefore seems to be the reason for larger $T_c$ in G3 set. 
The compressibility factor  $C_f$, which signify the deviation from ideal gas is given by \cite{Lourenco}
\begin{equation}
\label{compressibilityfactor}
C_f=\frac{P_c V_c}{T_c}=\frac{P_c}{\rho_c T_c},
\end{equation}
can be calculated for the respective force parameters. The universal class corresponding to the liquid-gas system have the compressibility factor $C_f \approx$ 0.292  \cite{cfuniversal} following the law of corresponding states. Table \ref{criticaldata} also shows the value of $C_f$ for each force parameter. All the forces except NL2 and NL-SH,  which have high incompressibility have $C_f$ close to 0.292.  This further validates the importance of an EoS well within the acceptable incompressibility range \cite{Lourenco}. \\
The recent parameters FSUGarnet, IOPB-I and G3 are known to work remarkably at zero temperature high-density regime.
This work also aims to find their validity at the low dense finite temperature limit.  For this,  EoS for each force in Fig. \ref{pvd} is also compared with the two-nucleon and three-nucleon interactions called $\nu_{14}$ and TNI \cite{fp} at T= 0, 5 and 20 MeV. This microscopic variational calculation is known to work better at low density but is not quite suitable for infinite nuclear matter \cite{variationalfail}. 
Whereas, here we are looking at the low-density behaviour of these forces at finite temperature. The EoS for each of the force at the mentioned temperature traces the microscopic calculation at density $<$ 0.06 MeV but then deviate at higher density.  This deviation gets better at a higher temperature and the parameter set G3 is nearest to the microscopic variational calculation at T=20 MeV. This is simply because the fitting procedure of any force does not take into account the critical temperature of the liquid-gas phase transition. The critical temperature, therefore, can be an important tool in making a force universally acceptable at both low and high-density regime at zero and finite temperature. It can be used along with the other bulk matter properties at zero temperature as a constraint for EoS. The parameter set G3 can be a good candidate for that. \\
\begin{figure}
	\centering
		\includegraphics[width=10cm,height=6cm]{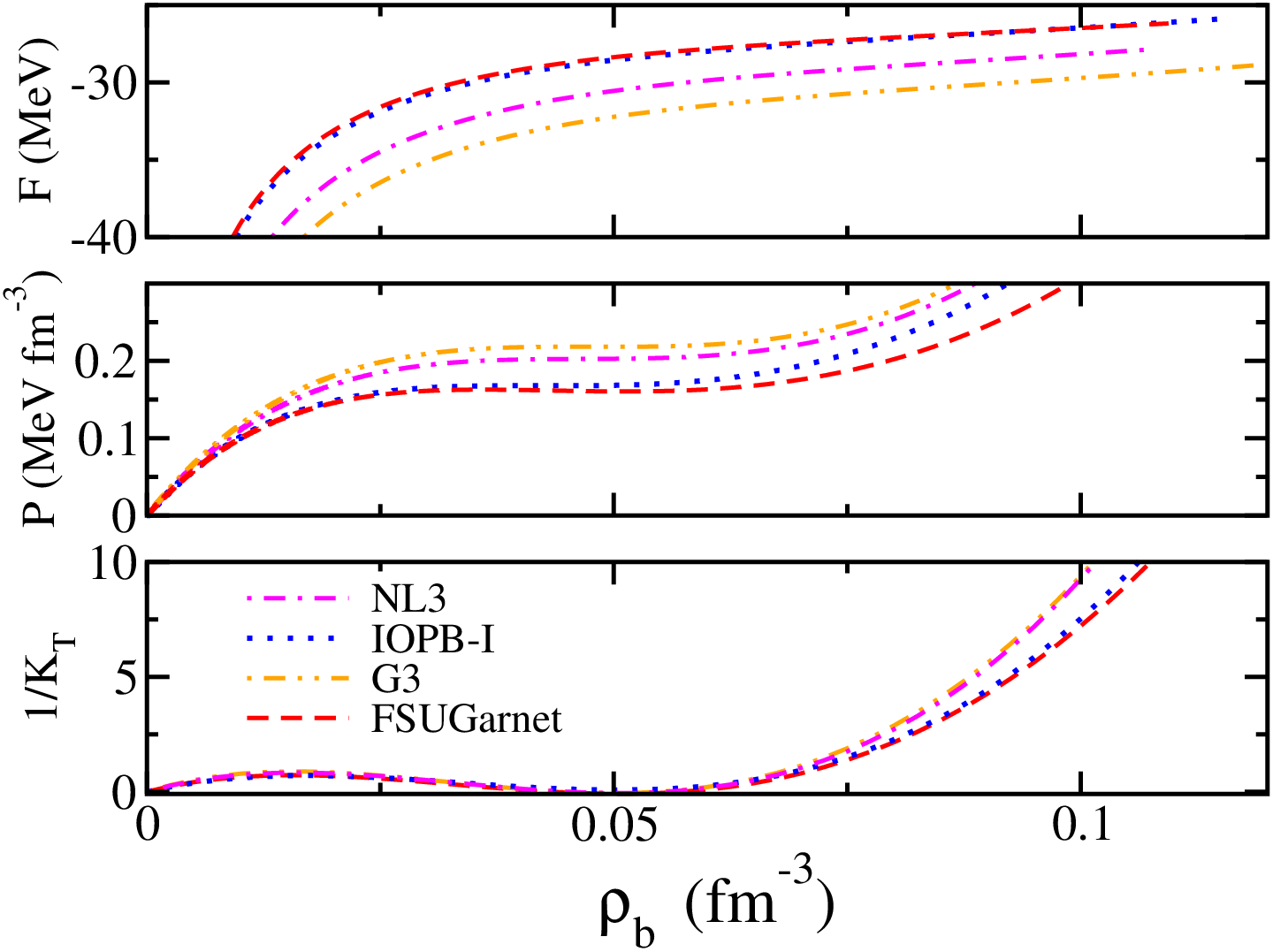}
	\caption{Variation of free energy, pressure and rigidity (1/K$_T$) with density for critical isotherm for the FSUGarnet, G3, IOPB-I and NL3 sets.}
	\label{rigidity}
\end{figure}
The temperature also impacts the incompressibility at saturation ($K_\infty$) or rigidity ($K_T^{-1} = \rho_b \partial P/ \partial \rho_b$) of nuclear matter. At large temperature, the nucleons are more prone to be free due to extra thermal energy and hence incompressibility decreases.
Figure \ref{rigidity} shows the variation of Free energy, pressure and rigidity (1/K$_T$) with density for critical isotherm for different parameter sets. For critical isotherm, the free energy is a smooth monotonically increasing function with G3   showing the least slope. Below the critical isotherm, the free energy has a well-defined pocket which disappears at T=T$_c$. The respective rigidity (1/K$_T$) is shown for each parameter set. It first increases and then starts decreasing and at the critical density, becomes zero. At critical density, one observes the inflation point in P-$\rho$ isotherm, which makes rigidity equals zero at the critical point.\\

\begin{figure}
	\centering
		\includegraphics[width=10cm,height=6cm]{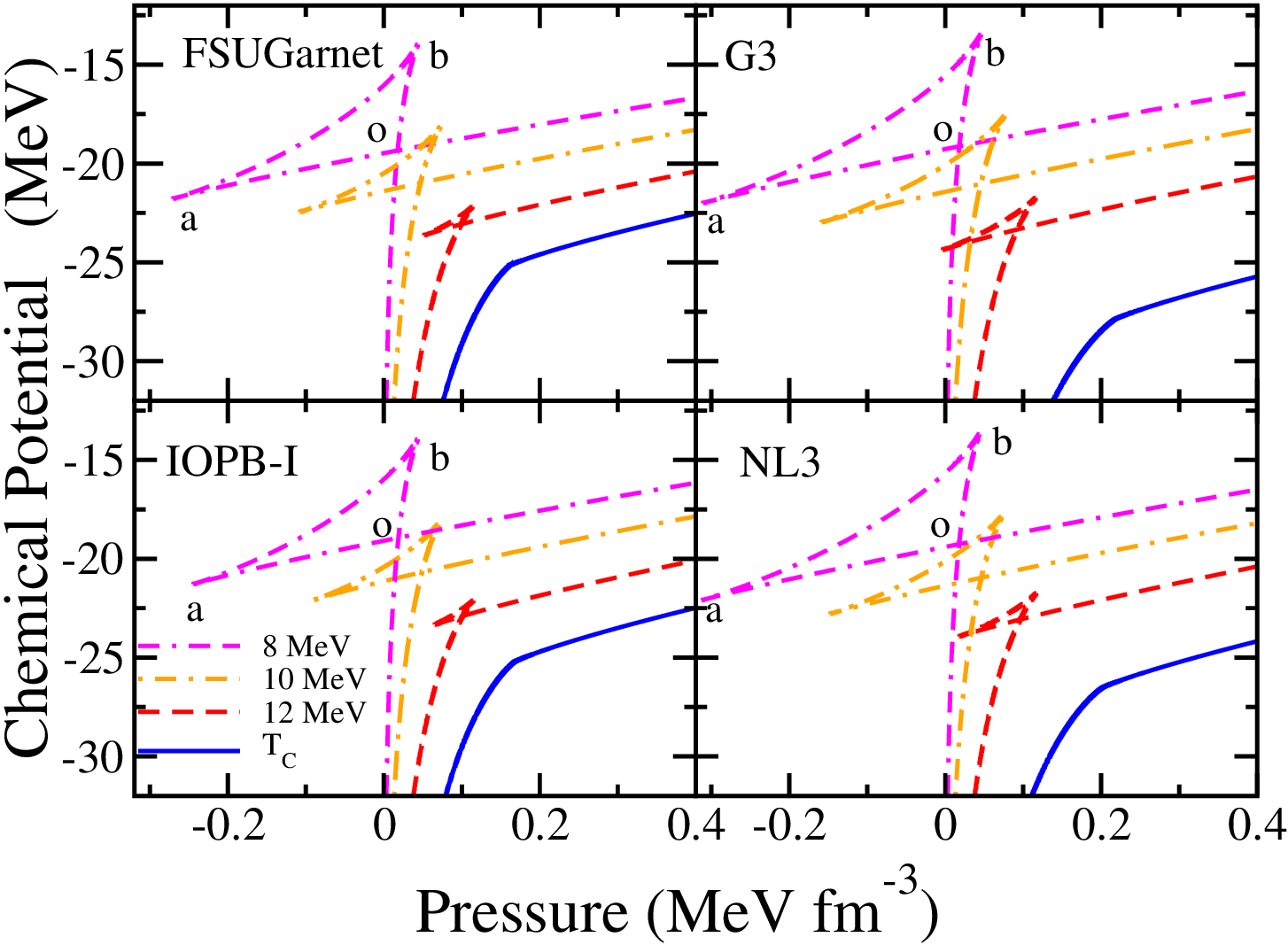}
	\caption{Chemical potential as a function of pressure for various values of temperatures.}
	\label{chemicalpotentialpressure}
\end{figure}
\begin{figure}
	\begin{center}
			\includegraphics[width=10cm,height=6cm]{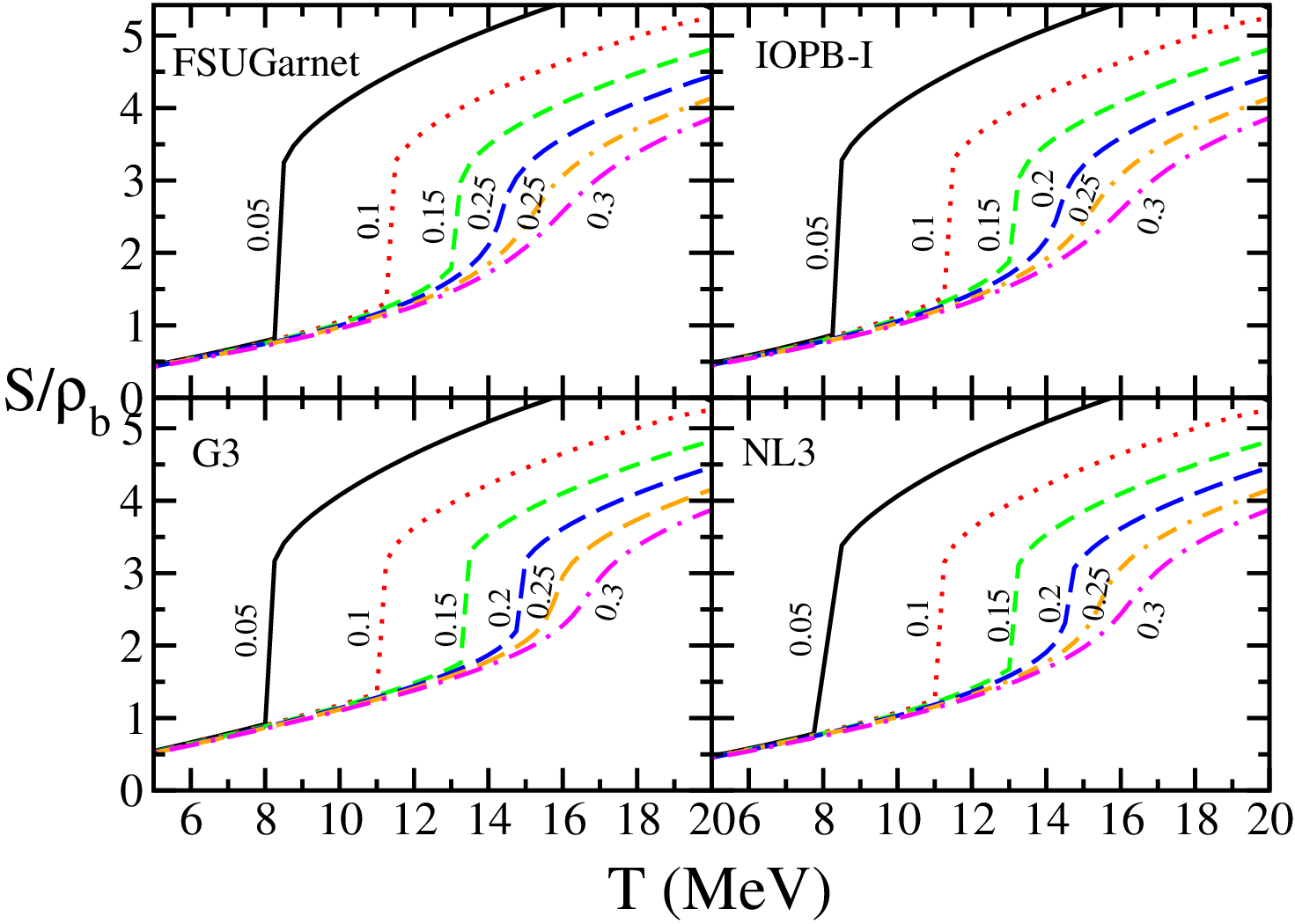}
		\caption{The entropy at constant pressure for 0.05, 0.1, 0.15, 0.2, 0.25 and  0.3 MeV fm$^{-3}$ as a function of temperature for various force parameters.}
		\label{isobarentropy}
	\end{center}
\end{figure}

The order of the transition is depicted in Fig \ref{chemicalpotentialpressure} in the chemical potential vs pressure graph.  The chemical potential plays an important role in determining the values of Fermi function which consequently determine the baryon density. For each parameter set, as one moves from zero pressure to point `b', chemical potential increases sharply. The potential then decreases till point `a' and pressure becomes negative. The path `ba' marks the instability region. After point `a', the potential increases monotonically with a different slope as compared to the path 0 to `b'. The area of path `obao' is largest for the set G3 and NL3 for each temperature. Because the system wants to attain minimum potential and free energy, a real system will not follow the path `obao' and will take the path with the least potential. It will move from zero pressure to point `o' and then will again increase. The path `obao' keeps decreasing with an increase in temperature and vanishes as one reaches the critical temperature. At critical isotherm, the slop changes discontinuously validating the argument of nuclear matter undergoing a first-order liquid-gas phase transition at low density.  The set G3 and NL3 estimate the largest instability area for a given isotherm, suggesting the dominance of strong attractive $\Phi$ field over repulsive vector $W$ field at low density for these sets, which play important role in determining the instability boundary. This is directly the consequence of the value of $\zeta_0, k_3 ,k_4$ in these forces.
The low value of $\zeta_0$ in G3 and NL3 sets account for the less repulsive force as compared to the high value of $\zeta_0$ in FSUGarnet and IOPB-I forces. Therefore the instability area is inversely proportional to the value of $\zeta_0$ in finite temperature limit. \\
The isobaric entropy as a function of temperature is shown in Figure \ref{isobarentropy} for the FSUGarnet, G3, IOPB-I and NL3 sets. The values of pressure are 0.05, 0.1, 0.15, 0.2, 0.25 and  0.3 MeV fm$^{-3}$ from left to right.  There is a sharp discontinuity at the transition temperature marking the presence of two phases in the system. As one goes beyond the critical temperature (refer Table \ref{criticaldata}), the curve becomes smooth and continuous showing that system now only exists in the gaseous  phase. One can see a sharp difference in isobar for 0.2 MeV fm$^{-3}$ for all the sets. For IOPB-I and FSUGarnet, this pressure is above $P_c$ showing continuous curve and for G3 and NL3, this is very close to $P_c$  thus still showing a discontinuity. Furthermore, the relative separation between isobars decreases as we increase the pressure.\par
The isobaric entropy is a function of chemical potential as is shown in Eq. \ref{entropyeq} and consequently dictated by the respective EoS. Again the vector self-coupling $\zeta_0$ plays an important role. It can, therefore, be apprehended that, if a force parameter is designed to have the $\zeta_0$ $\approx$ 1.0  keeping the incompressibility within the limit, it can address the qualitative as well as quantitative nature of phase transition. This is visible in these parameter sets. They only differ much in the value of $\zeta_0$ with almost the same bulk matter properties. The force G3 stands out due to its low $\zeta_0$ and pressence  cross-couplingss ($\eta_1, \eta_2$) of scalar $\sigma$ and vector $\omega$ meson.  

\subsection{\label{StabilityAnalysis} Stability Analysis}
Let us consider the symmetric nuclear matter interacting only through strong interaction as Coulomb interaction is zero hypothetically due to absence of any source current \cite{zerocoloumb}. It, therefore, reduces to a one-component system and Gibbs phase rule then allows only one degree of freedom i.e. only one pressure for each temperature. The phase transition of nuclear matter liquid into gaseous phase start occurring when the stability condition \cite{pwang}
\begin{equation}
F(T,\rho) < (1-\lambda) F(T,\rho')+\lambda F(T,\rho''),
\end{equation}
with
\begin{equation}
\rho= (1-\lambda)\rho'+\lambda \rho'', \hspace{0.3cm} (0<\lambda<1),
\end{equation}
where F is Helmholtz free energy per volume and  $\lambda$ is volume fraction,   is violated. 

The two phases are denoted by a prime and a double prime. This implies that a system should follow the inequalities
%\begin{eqnarray}
\begin{align}
C_V >0 \; or\; \Big[\frac{\partial S}{\partial T}\Big]_\rho >0,\\
\label{ineq}
K_T>0\; or  \Big[\frac{\partial P}{\partial \rho}\Big]_T >0,
\end{align}
%\end{eqnarray}
which corresponds to the dynamical and mechanical stability, to prevent phase change. If anyone of these is violated, a system with more than one phase is energetically favourable. 
In our calculations, the dynamical stability condition is never violated. 
In the limit $T \rightarrow 0$, the specific heat vanishes as $dS/dT \rightarrow 0$. At higher temperature, the specific heat asymptotically approaches the noninteracting limit 3/2 for low densities.

\begin{figure}
	\centering
		\includegraphics[width=10cm,height=6cm]{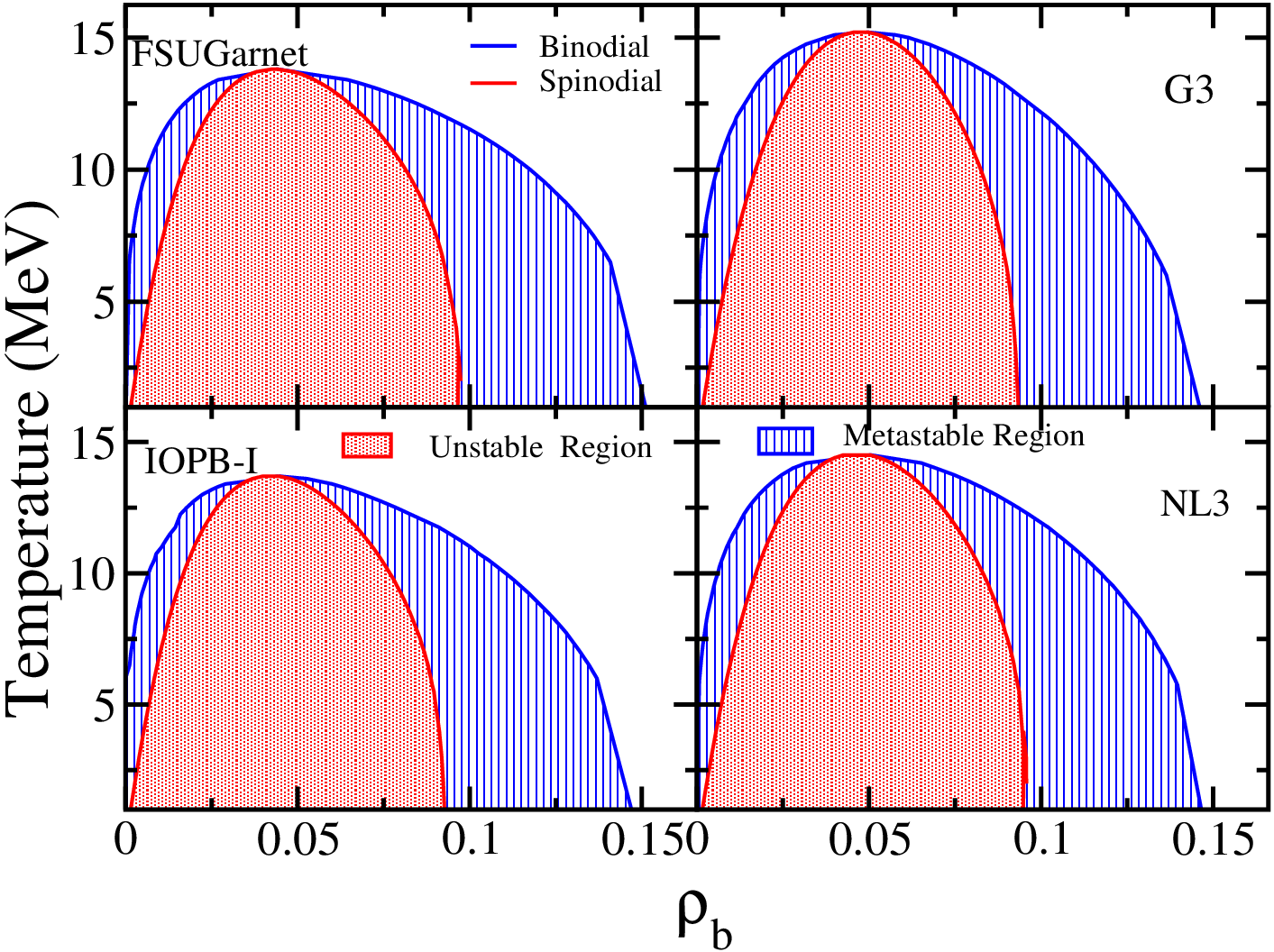}
	\caption{The binodal and spinodal curve for FSUGarnet, G3, IOPB-I and NL3 sets .}
	\label{binodal}
\end{figure}

In Fig. \ref{pvd}, one can observe that for each isotherm, there are three values of densities for each small positive pressure for $T< T_c $. At very low density, the thermal and Fermi degeneracy pressure makes the total pressure positive. As density increases slightly, the attractive force arising due to attractive scalar $\Phi$ field, increases, which try to reduce the volume of the system and pressure goes from positive to negative. This corresponds to the part where $dP/d\rho < 0$. It violates the mechanical stability inequality in Eq. \ref{ineq} and system transform into a two-phase system. Instability is only a sufficient condition for phase separation to take place, therefore, it may even occur in the one phase-stable system. This is the case of the metastable state \cite{metastablestate}. Once the two-phase system is favoured over the one-phase system, the solution of Eqns. \ref{chmeicalpotentialeqi} and \ref{pressureeqi} gives us the coexistence densities i.e. $\rho_g$ for gaseous phase and $\rho_l$ for liquid phase for each isotherm. This density pair for each temperature and pressure then determine the coexistence phase boundary or binodal. Also, the locus of points where the second derivative of free energy is zero for each isotherm marks the boundary of spinodal. The point where the binodal and spinodal curve meet defines the critical parameters given in Table \ref{criticaldata}. This is shown in Figure \ref{binodal} where the binodal or coexistence boundary and spinodal or instability boundary is plotted for FSUGarnet, G3, IOPB-I and NL3 sets. The middle red shaded region marks the region of instability and outer blue region signify the metastable region. The outer blue solid line is the coexistence boundary which gives us the ($\rho_g, \rho_l$) couple.

\begin{figure}
	\centering
		\includegraphics[width=12cm,height=6cm]{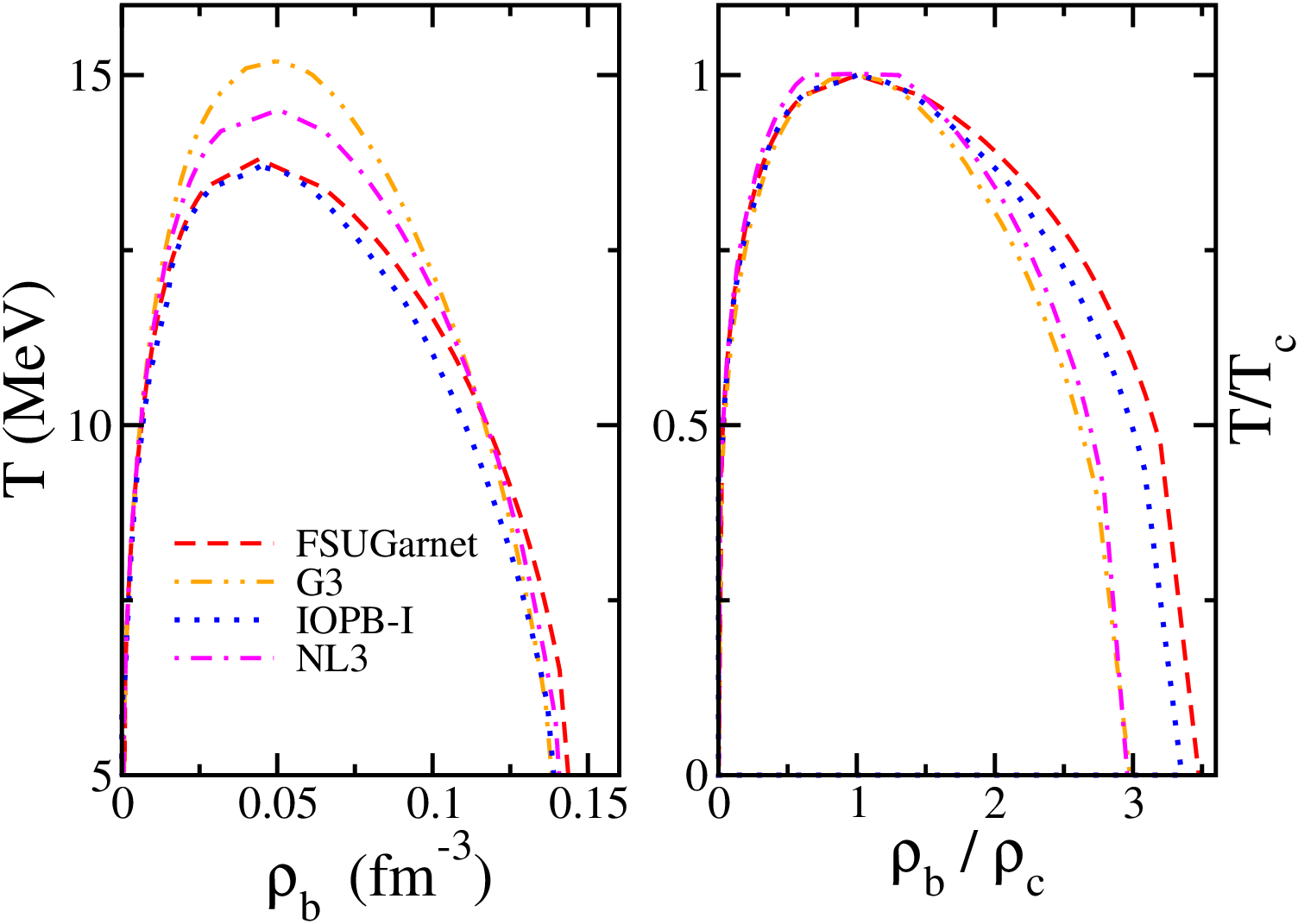}
	\caption{The binodal curve for FSUgarnet, G3, IOPB-I and NL3 sets  along with dimensionless density and temperature}
	\label{binodal1}
\end{figure}

\begin{figure}[H]
	\centering
		\includegraphics[width=12cm,height=6cm]{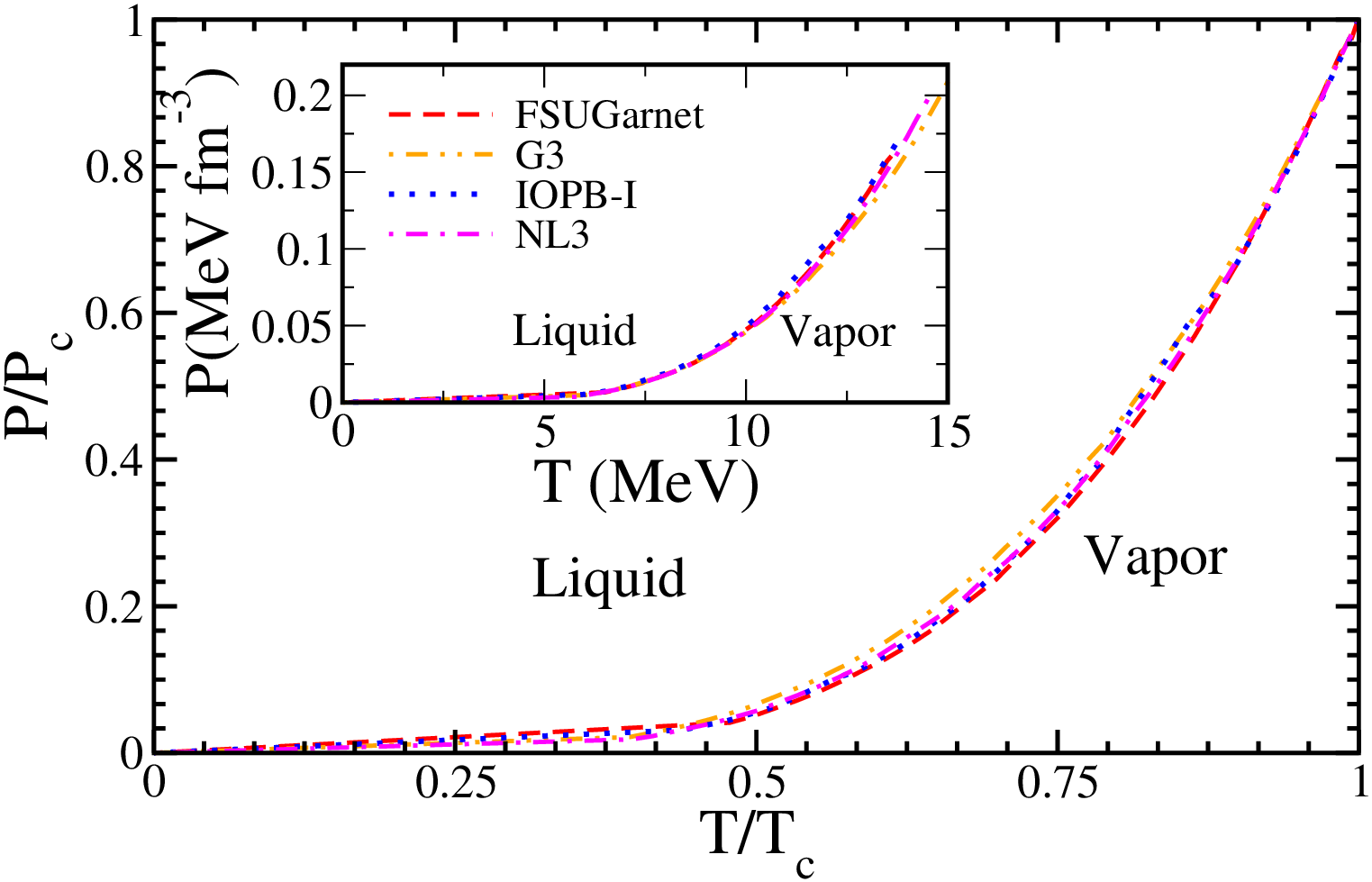}
	\caption{Vapour pressure as a function of temperature with absolute and reduced dimensionless values.}
	\label{ptcurve}
\end{figure}

The liquid at saturation density is in equilibrium with the zero density gaseous phase at zero temperature. As one increases the temperature, the liquid coexistence density decreases and gaseous coexistence density increases until the critical temperature is encountered. Beyond critical temperature, the liquid-gas coexistence vanishes and the system only remain in the gaseous phase. At small temperatures i.e. below 5-6 MeV, there is no solution to Eqns. \ref{chmeicalpotentialeqi} and \ref{pressureeqi} and system only remain in one stable liquid phase. For more insight and closer comparison among all the force parameters, the binodal are plotted in Figure \ref{binodal1}. Here, critical density and temperature are used for scaling in the right panel. As discussed, the least value of $\zeta_0$ gives us the maximum binodal area for G3. The NL3 shows a similar behaviour. on the other hand, FSUgarnet and IOPB-I  with comparable $\zeta_0$ show similar behaviour if we look at the right panel of the Figure \ref{binodal1}. The comparatively large instability region in G3 is the consequence of soft scalar-self coupling as also evident in graph (c) of Figure \ref{comparison}.\\
One of the unique features of the first-order phase transition is that the first-order derivatives of Gibbs free energy are discontinuous. It gives rise to the release of heat at a constant temperature which is called the latent heat of vaporization. The latent heat can be determined using the Clausius Clapeyron equation given in Eq    \ref{classiousclapron}. This equation requires the derivative of pressure with temperature along the binodal or coexistence curve.  This is called vapour pressure. Figure \ref{ptcurve} shows the variation of vapour pressure along with the temperature for the NL3, FSUGarnet, IOPB-I, and G3 sets. The outer graph is in the reduced dimensionless form where the respective critical parameter is used to making them dimensionless. The inner graph uses absolute values of the nuclear matter parameters. There is not much deflection in different force parameter sets. At higher temperatures, the curve has a linear dependence which can be expressed as $\pi = a \tau$ where $\pi$ and $\tau$ are reduced pressure and temperature respectively. This characteristic helps to determine the behaviour of the curve in the neighbourhood of critical parameters. At lower temperature, pressure rises very slowly and there is a sudden increase after 0.5$T_c$ for all forces. Below this temperature system only remains in one stable liquid phase and therefore system encounter very less pressure. At higher temperature, the thermal excitation and existence of a two-phase system drive the pressure towards higher values.\\

\begin{figure}[H]
	\begin{center}
			\includegraphics[width=10cm,height=6cm]{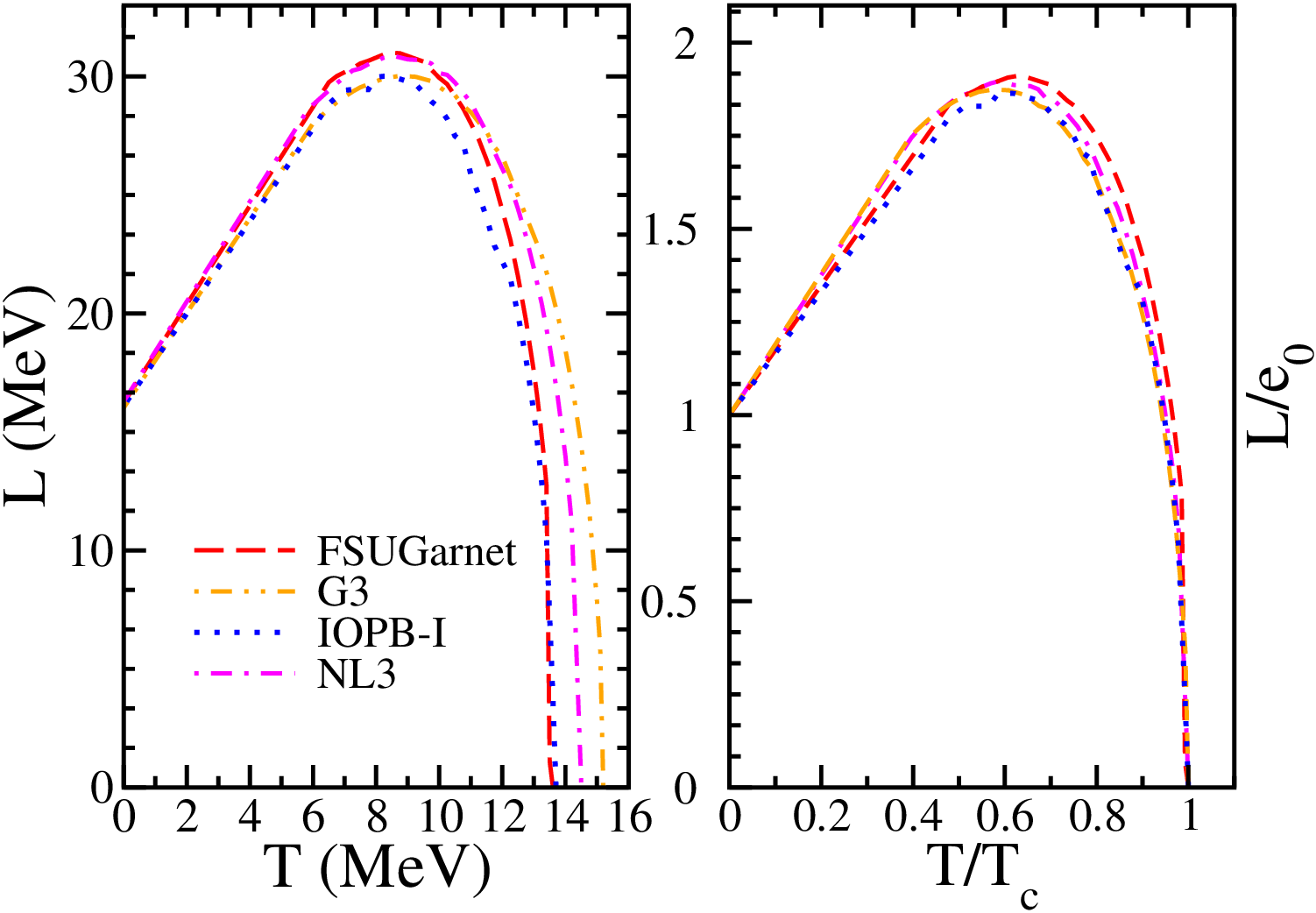}
		\caption{Latent heat of vaporization for symmetric nuclear matter $L=T(s_g-s_l)$ for the FSUGarnet, IOPB-I, NL3 and G3 sets}
		\label{lheat}
	\end{center}
\end{figure}
Figure \ref{lheat} shows the variation of latent heat of vaporization with temperature for different parameter sets using the relation $L=T(s_g-s_l)$ by virtue of Clausius Clapeyron equation. This method is numerically more stable as compared to the calculation of the derivative of the vapour pressure shown in Figure \ref{ptcurve}. We have checked the values of latent heat with both the methods and they estimate a similar trend. The figure shows the characteristic behaviour of latent heat with temperature. In the zero-temperature limit, we can write the Clausius Clapeyron equation as \cite{0templheateq}
\begin{equation}
\small
\lim_ {T \to 0 } L(T) = \lim_{T \to 0} T\Big(\frac{1}{\rho_g}-\frac{1}{\rho_l}\Big) \frac{dP_{vapour}}{dT}= -\mu_g = -\mu_l= e_0,
\end{equation}
because the liquid-gas coexistence density approaches zero for vanishing temperature. Therefore, in this limit, latent heat is simply the heat required to remove a nucleon from the saturated liquid with energy density e$_0$. As one increases the temperature, the latent heat rises linearly up to a maximum value which is close to $\approx$ 30 MeV. This linearity at low temperature indicates the absence of interactions for low-density gas. This trend is similar to all the forces considered and there is very less deflection among different parameters in both left panel where absolute values are plotted and in the right panel where reduced parameters are depicted to minimise the mean-field dependence i.e. effect of different vector and scalar self and cross-couplings of the various models. The latent heat reaches the peak value in between 8-10 MeV or (0.5-0.7)T$_c$ in all cases and then fall sharply to zero at the critical temperature. The maximum $L/e_0$ in right panel has a very narrow range of 1.8-1.9. The Latent heat, therefore, is a thermally correlated parameter. Furthermore, the maximum latent heat L$_H$  is the least correlated parameter at finite temperature. A similar trend is seen if we use flash temperature as the scaling parameter. Therefore, flash temperature along with the critical temperature is suitable to constraint the EoS for finite temperature. At low temperature, the latent heat has linear relationship with temperature and  as the system reaches near the critical temperature, all the force parameters give the same slope in the right panel of the graph which is a significant validation of the well defined thermodynamical theories \cite{0templheateq}.

\subsection{\label{scaling}Scaling Laws}
The critical points related to the phase transition are of special interest as they are used to subsequent characterisation.     
The behaviour of a given system in the neighbourhood of these points forms a fundamental problem in view of the phase transition phenomenon. Near the critical point, various physical quantities encounter the singularity. It becomes essential to express these singularities in terms of power laws.
These power laws are characterised by critical exponents and in turn, determine the qualitative nature of a given system.  The exponent should also follow well known scaling laws which make only two exponents independent at a given time. The nature of exponents is such that they do not differ much from one system to other in one universality due to their dependence on very less number of parameters \cite{patheria}. The E-RMF calculations are needed to numerically satisfy these laws to check their independence from the underlying interactions. Therefore, these are used here to   qualitatively check the consistency of our analysis.
The exponents $\beta$, $\lambda$, $\delta$, $\alpha$ and $\tau$ are defined as \cite{jbeliot}
\begin{eqnarray}
\rho_l-\rho_g &\sim & \Big(\frac{T_c-T}{T_c}\Big)^\beta, \\
K_T &\sim &\Big| \frac{T-T_c}{T_c}\Big|^{-\lambda},\\
|P-P_c|_{T_c}& \sim &\Big|\frac{\rho-\rho_c}{\rho_c}\Big|^\delta ,\\
C_V & \sim &(T_c-T)^{-\alpha},
\end{eqnarray}
and finally, from Fisher's analysis \cite{fisher},
\begin{eqnarray}
C_f= \frac{\zeta(\tau)}{\zeta(\tau-1)}.
\end{eqnarray}
A simple power law fitting was done in the form $y=aX^b$ and corresponding values of these exponents along with the mean field reaults and other liquid-gas systems are shown in Table \ref{criticalexponent}. All the critical exponents are close to their respective values derived from mean field results.
\begin{table}
	\centering
	\caption{Calculated critical exponent for NL3, IOPB-I, FSUGarnet and G3 parameter sets. }
	\begin{tabular*}{\linewidth}{c @{\extracolsep{\fill}} ccccccc}
		\hline
		\hline
		Critical   & NL3 & IOPB-I & FSUGarnet & G3 & Mean  & Liquid-Gas   \\ Exponent&&&&&Field&System \cite{gasliquidsystem}\\
		\hline
		$\beta$ &0.37 &0.40&0.39  &0.40  &0.5  &0.32-0.35  \\ 
		$\lambda$ &0.97 &1.12  &1.02&1.01  & 1 &1.2-1.3  \\ 
		$\delta$ & 3.65 &3.62 &3.50  &3.38 & 3 &4.6-5.0  \\ 
		$\alpha$ & 0.00 &0.00 &0.00 &0.00& 0 &0.0-0.2  \\ 
		$\tau$ &2.209 & 2.210 &2.167 &2.224&- &2.1-2.25  \\ 
		\hline  
		\hline  
	\end{tabular*}
	
	\label{criticalexponent}
\end{table}
The consistency check of the critical exponents can be done using standard scaling laws or thermodynamic inequalities \cite{scalinglaw}. These are given by Rushbrooke, Griffiths and Widom inequalities and can be respectively written as \cite{jbeliot} 
\begin{eqnarray}
\alpha + 2 \beta + \gamma \ge 2,\\
\alpha +  \beta(1+ \delta) \ge 2,\\
\beta(\delta-1) - \gamma \le 0,
\end{eqnarray}
and from Fisher's droplet model as ,
\begin{eqnarray}
\frac{\beta}{\gamma}- \frac{\tau-2}{3-\tau} = 0.
\end{eqnarray}

Table \ref{scalinglaw} compiles all the calculated results of these scaling laws
for each parameter set. The critical exponents derived from each parameter satisfy these scaling laws up to a good approximation. We, can therefore strongly conclude that E-RMF parameter sets with incompressibility comparable to accepted experimental value are very consistent in the finite temperature limit satisfying the statistical inequalities. They can consequently be used in the calculation of hot EoS which is of primary importance in processes like binary neutron star merger where temperature plays a very important role.   
\begin{table}[ht]
	\centering
	\caption{Result of scaling laws for NL3, IOPB-I, FSUGarnet and G3 parameter sets.}
	\begin{tabular*}{\linewidth}{c @{\extracolsep{\fill}}  ccccc}
		\hline
		\hline
		Force Parameter & Rushbrooke  & Griffiths & Widom & Fisher \\
		\hline    
		
		NL3&1.72&1.74  &0.02& 0.13 \\
		IOPB-I&1.92  & 1.84 &-0.07&0.09  \\
		FSUGarnet& 1.79  & 1.75 &-0.04&0.10   \\
		G3&1.80  &1.75  & -0.05&0.10 \\
		\hline     
		\hline     
	\end{tabular*}
	
	\label{scalinglaw}
\end{table}

\section{\label{summary} Summary and Conclusions}

In summary, we have discussed that the nuclear matter dictated by strong interactions which have an attractive (scalar)  and a repulsive (vector ) part undergo a first-order liquid-gas phase transition at low densities alike Van der Waals equation of state. We have used three E-RMF force FSUGarnet, IOPB-I and G3 in the finite temperature limit realizing there narrow range of bulk matter properties at zero temperature. They are subsequently compared with one of the most used set NL3. They all have different fitting procedures and therefore different coupling constants. This fact is utilized to understand the role of these couplings in the liquid-gas phase transition of nuclear matter. It is shown that these forces can describe this phase transition in an excellent manner qualitatively using instability analysis and scaling laws. In our calculations, the critical temperature for the symmetric nuclear matter lies between (13.75-15.3) MeV. The corresponding range of pressure $P_c$ and density  $\rho_c$ are between (0.167-0.218) MeV fm$^{-3}$ and (0.042-0.049)fm$^{-3}$, respectively. The estimated critical parameters are in agreement with other RMF forces which have incompressibility in the range 240 $\pm$ 20 MeV.\\
The critical parameters i.e. $T_c$, $P_c$ and $\rho_c$ are clearly model dependent. However, there is a direct correlation among  parameter such as $\rho_c\rho_f$, $T_cT_f$, $P_cT_c$, etc. The critical temperature is not constrained in theoretical as well as in experiments. Consequently, there is large uncertainty in the value of critical temperature among experiments as well as theoretical calculations with same bulk matter properties. Both the experimental and theoretical calculations of critical temperature are model dependent. The value of critical temperature $T_c$, therefore, can be used for constraining the EoS.\\
The vector-self coupling is a determining factor in these E-RMF forces. Force with a lower value of $\zeta_0$ estimate large coexistence area and large $T_c$. Similarly, $T_c$ is positive correlated with incompressibility at saturation except in NL3 set. A large negative value of $k_4$ makes the NL3 behave differently in some cases as compared to other forces. 
The parameter sets used in the present work estimate the compressibility factor (C$_f$) close to the universal value of liquid-gas systems. Whereas, the parameters having higher incompressibility deviate much from this value. We have also checked the consistency of our calculation using critical exponents and scaling laws and emphasize that E-RMF parameters sets with acceptable incompressibility are consistent and reliable at finite temperature limit. All the critical exponents are very close to the mean-field results and experimental liquid-gas systems.\\
We have also studied the stability conditions in terms of binodal and spinodal curves using IOPB-I, FSUGarnet, G3 and NL3 force parameters. Subsequently, they are used to determine thermodynamic properties like specific and latent heat. The specific heat has linear dependence for a large number of particles. For a lesser number of particles, it asymptotically approaches the classical limit. Latent heat, on the other hand, is found to be a thermally correlated variable and does not depend on the effective field.\\
The E-RMF sets explain the phase transition in the low-density regime exceptionally well. A little discrepancy in the experimental and theoretical values might be attributed to two main factors: (1) symmetric nuclear matter being an ideal system which is too difficult to simulate in experiments and (2) finite size effect and very short time scale of multi fragment reactions which make it difficult to study thermodynamic equilibrium.  Moreover,  there is also a need to look for higher-order vector interaction terms which might influence the EoS at finite temperature keeping other properties intact.

\end{document}